\documentclass[journal=jacsat,manuscript=article,layout=twocolumn]{achemso}
\usepackage{bm}
\usepackage{amsmath}
\usepackage{braket}
\usepackage{pdfpages}

\def\[#1\]{\begin{align}#1\end{align}}
\newcommand{\degree}{^{\circ}}
\newcommand{\AAA}{\text{\AA}}

\title[Ice-like Surface]{Local ice-like structure at the liquid water surface}

\author{Nathan L. Odendahl}
\affiliation{Department of Chemistry, University of California, Berkeley, California 94720, USA}
\alsoaffiliation{Chemical Sciences Division, Lawrence Berkeley National Laboratory, Berkeley, California 94720, USA}
\author{Phillip L. Geissler}
\affiliation{Department of Chemistry, University of California, Berkeley, California 94720, USA}
\alsoaffiliation{Chemical Sciences Division, Lawrence Berkeley National Laboratory, Berkeley, California 94720, USA}
\email{geissler@berkeley.edu}

\begin{document}

\section*{Abstract}

Experiments and computer simulations have established that liquid water's surfaces can deviate in important ways from familiar bulk behavior. Even in the simplest case of an air-water interface, distinctive layering, orientational biases, and hydrogen bond arrangements have been reported, but an overarching picture of their origins and relationships has been incomplete. Here we show that a broad set of such observations can be understood through an analogy with the basal face of crystalline ice.
Using simulations, we specifically demonstrate that water and ice surfaces share a set of structural features
suggesting the presence of nanometer-scale ice-like domains at the air-water interface. Most prominent is a shared characteristic layering of molecular density and orientation perpendicular to the interface.
Similarities in two-point correlations of hydrogen bond network geometry point to shared ice-like intermolecular structure in the parallel direction as well.
Our results bolster and significantly extend previous conceptions of ice-like structure at the liquid's boundary, and suggest that the much-discussed quasi-liquid layer on ice evolves subtly above the melting point into a quasi-ice layer at the surface of liquid water.

\section{Introduction}

The boundary of a macroscopic liquid -- whether at an electrode, a macromolecular surface, or an interface with a coexisting phase -- breaks symmetries and imposes constraints that can alter microscopic structure and response in important ways.
In the case of water, changes in the statistics of intermolecular arrangements have been discussed as key factors in, e.g., surface chemistry\cite{Ocampo1982,Kang2005,PinheiroMoreira2013,Bartels-Rausch2014,Gerber2015,Hudait2018} atmospheric aerosol behavior\cite{Gerber2015,Laskin2003,Finlayson-Pitts2010}, and ice nucleation\cite{Ocampo1982,Hudait2018,Haynes1992,Cox2015,Bi2016,Qiu2017,Hudait2019}.
The detailed molecular physical origin of surface effects, however, remains a subject of uncertainty and debate in many of these areas\cite{Shin2018,Shin-Willard,Serva2018,Besford2018,Kessler2015,Bjorneholm2016}.

The subtlety of such interfacial phenomena arises in part from the inherently transient and short-ranged character of liquid structure.
In the bulk liquid phase, microscopic structure is apparent only in those observables that break translational symmetry, either by referencing multiple points in space or/and by tagging a specific molecule.
An extended surface breaks translational symmetry along the perpendicular Cartesian coordinate $z$, so that structure is manifest in even single point observables,
such as the density profile $\rho(z) = (N/A)\langle \delta(z-z_1)\rangle$ where $N$ is the number of molecules, $A$ is the interface's area, and $z_1$ is the vertical position of a particular molecule.
But for soft aqueous interfaces, such as the air-water boundary, profiles of this kind are typically nondescript -- a smooth crossover from the density of one phase to the other, or a slight net molecular orientation decaying rapidly into the bulk phases.
It is difficult to draw or defend any detailed inference from such measurements, even with the nearly unlimited resolution of molecular simulation.

Experimentally, microscopic characterization of liquid-vapor interfaces is hindered further by the complicated nature of surface-specific spectroscopies. Vibrational sum-frequency generation (VSFG) reports specifically on molecular environments that lack centrosymmetry, and is therefore well suited to study interfacial structure that mediates frequencies of bond vibration. 
Mapping VSFG spectra back onto intermolecular structure, however, typically requires uncontrolled approximations and the assistance of molecular simulation.  
Even then, conflicting conclusions have been drawn for some systems. 
Early VSFG measurements of water's surface\cite{Du1993,Du1994} revealed a strong signal near 3700 cm$^{-1}$ that is broadly agreed to originate in hydroxyl vibration of water molecules with H atoms exposed to the vapor phase, often termed the ``dangling'' or ``free'' OH group.
Spectral features in the range 3100-3500 cm$^{-1}$, consistent with hydroxyl stretching of intact hydrogen bonds, have been variously interpreted in terms of coupling among vibrational modes within and among molecules\cite{Raymond2003,Sovago2008,Auer2008,Yang2010,Stiopkin2011,Ishiyama2012,Medders2016},
or in terms of discrete populations of distinct hydrogen bonding
environments\cite{Buch2005,Walker2006,Ji2008,Nihonyanagi2009,Pieniazek2011,Stiopkin2011,Byrnes2011,Ishiyama2012,Sun2018,Smolentsev2017,Tang2020}.

Several results from these observations and simulations of air-water interfaces have inspired analogies with the crystalline phase.
The aforementioned 3700 cm$^{-1}$ VSFG peak, commonly attributed to dangling OH groups, is observed as well for the interface between vapor and crystalline ice. Indeed, dangling OH groups are a characteristic feature of ice's low-energy basal plane.
In addition, phase-sensitive VSFG measurements exhibit a pronounced band around 3200 cm$^{-1}$, consistent with OH stretching in bulk ice\cite{Ji2008}.
Based on these similarities, Shen and Ostroverkhov\cite{Shen2006} suggested that ice-like structure is a characteristic feature of the liquid's boundary.
Fan et al.\cite{Fan2009} supported this connection by computing depth-resolved probability distributions of molecular orientation from molecular dynamics simulations of the air-water interface.
They observed a clear anisotropy within two molecular diameters of the average interfacial height, with layered features that are reminiscent of ice's basal facet but are highly broadened and only weakly distinct from one another.
These results support a loose structural resemblance between air-water and air-ice interfaces.

As noted in Fan et al.\cite{Fan2009}, nondescript structural profiles like $\rho(z)$ are to be expected at soft interfaces, which are highly permissive of capillary wave-like fluctuations in interfacial topography.
By contrast, liquid structure at hard interfaces can exhibit pronounced spatial variations that extend several molecular diameters into the liquid phase.
In computer simulations of water confined between flat hydrophobic plates, Lee et al.\cite{Lee1984} revealed substantial oscillations in average microscopic density that decay on a nanometer length scale; distributions of molecular orientation are also strongly peaked near the interface.
A detailed comparison between the liquid's surface and the crystalline structure of ice is thus more straightforward in this case of a hard, flat, hydrophobic interface. 
Lee et al. found a close correspondence: Spacing between peaks of $\rho(z)$ suggest a significant deviation from bulk liquid structure, aligning more closely with the separation between layers of ice in the direction normal to the basal face.
Similarly, preferred molecular orientations at the peaks of $\rho(z)$ are roughly consistent with hydrogen bonding directions in the corresponding ice layers, though with exceptions that suggest specific distortions of the basal plane.

More recent computational studies of soft aqueous interfaces have emphasized that the smooth density profile $\rho(z)$ belies a sharpness that is evident in any representative molecular configuration. 
For a given lateral position $(x,y)$ in such a configuration, the dense liquid environment gives way to dilute vapor over a very small range of $z$ centered at 
$z_{\rm inst}(x,y)$.
With the vertical coordinate referenced to this local ``instantaneous interface", Willard and Chandler demonstrated that the air-water interface is in fact highly structured\cite{Willard2010}, with average features even more pronounced than Lee et al. reported for water at a hard hydrophobic interface.
With this definition of depth (and related definitions\cite{Chacon2003,Sega2015,Partay2008}), distinct interfacial layers have been identified for a variety of microscopic properties, including density\cite{Willard2010,Sega2015,Kessler2015}, orientation\cite{Partay2008,Willard2010,Kessler2015,Wohlfahrt2020}, hydrogen bond formation\cite{Partay2008,Kessler2015,Serva2018}, bond network connectivity\cite{Partay2008,Zhou2017,Pezzotti2017, Pezzotti2018}, molecular dipole\cite{Shin-Willard}, and hydrogen bonding lifetime and bond libration\cite{Tong2016,Zhou2017}.

Models and physical pictures have been offered to rationalize these microscopic patterns at the air-water interface\cite{Lee1984,Wilson1988,Ismail2006,Fan2009,Bonthuis2011,Vaikuntanathan2016,Pezzotti2017,Shin2018,Shin-Willard}, which become apparent only after an accounting for its fluctuating surface topography.
But, to our knowledge, an overarching structural analogy with ice has not been evaluated from the instantaneous interface perspective.
This paper presents such an evaluation, based on molecular simulations, whose results strongly encourage the notion of ice-like organization at the air-water interface.
Specifically, we will show that most sharp features present in instantaneous depth profiles of molecular density and orientation can be anticipated from molecular arrangements at the basal face of an ideal ice lattice.
Where this correspondence fails, similar deviations are observed at the surface of ice at finite temperature (but still well below the melting point), due to low-energy defects that distort its ideal lattice structure.
This analogy highlights local structural motifs that are common to the surfaces of ice and water, but does not imply correlations in liquid structure that extend beyond a few molecular diameters.
Instead, we find that the accumulation of defects at the air-ice interface progressively degrade lateral correlations as temperature increases, producing distinctive features that are shared on both sides of the melting transition.

\subsection{Ice structure}
\begin{figure}[t]
	\includegraphics[width=\linewidth]{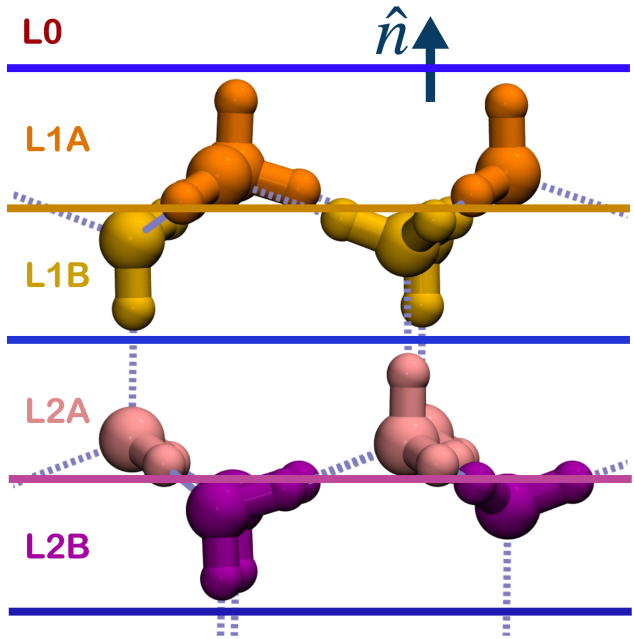}
	\caption{
	Ideal lattice structure at the basal plane of ice, with outward-pointing normal vector $\hat{\bf n}$.
	Colors highlight a characteristic layering of chair-configuration hexagonal motifs. Only the positions of molecular centers, and the connectivity of hydrogen bonds, is prescribed by this ideal structure. For concreteness we show one configuration of protons consistent with the Bernal-Fowler ice rules. The lattice continues indefinitely in the left, right, and downward directions; L0 marks the beginning of the vapor phase.
	}
	\label{fig:two-layer-schematic}
\end{figure}

The essential structure of ice has been known for nearly 100 years, since the extended tetrahedral network of hexagonal ice was proposed following the rules of Bernal and Fowler\cite{Bernal1933}. 
Linus Pauling argued that the fundamental building block of ice must be the ``puckered'' or ``chair''-configuration hexagon\cite{Pauling1935} (comprising, e.g., the orange and yellow molecules of Fig.~\ref{fig:two-layer-schematic} labeled L1A and L1B).
These puckered hexagons tessellate across a plane, forming a bilayer of slightly higher (orange in Fig.~\ref{fig:two-layer-schematic}) and slightly lower (yellow in Fig.~\ref{fig:two-layer-schematic}) waters that constitute the basal (0001) face of ice. 
Each water molecule, at a vertex of three hexagons, forms three hydrogen bonds within the bilayer and reaches either upward (toward the vapor) or else downward to form a fourth hydrogen bond with an adjacent hexagonal bilayer. 
Hexagonal ice can therefore be viewed as stacks of puckered hexagonal planes that are interconnected with ``pillars'' in the form of hydrogen bonds. 
We will denote these bilayers L1, L2, etc.
The upward- and downward-reaching molecules within layer L$X$ will be designated L$X$A and L$X$B, respectively (see Fig.~\ref{fig:two-layer-schematic}).

At temperatures well below freezing, macroscopic ice crystals tend to terminate with a full bilayer of the basal plane exposed\cite{Materer1997}. 
In a real system, this termination generates some degree of restructuring, but the unaltered ``ideal" ice surface serves as a useful reference for our work.
Because the outermost molecules (all those in L1A) are lacking one hydrogen bond, they are also the most likely participants in rearrangements at finite temperature.
As the melting point is approached, discrete defects accumulate at the ice surface\cite{Bishop2009,Buch2005,Buch2008,Smit2017B}, which develops a much-discussed\cite{Rosenberg2005,Li2007,Conde2008,Bishop2009,Watkins2011,Limmer2014,Michaelides2017,AlejandraSanchez2017,Pickering2018,Kling2018,Mohandesi2018,Slater2019} boundary region (called a pre-melted, quasi-liquid, or liquid-like layer) with fluid-like characteristics.
This region of 1-3 ice bilayers\cite{Conde2008,Kling2018} clearly mediates the contact of coexisting crystal and vapor phases, but the nature of its onset\cite{Kling2018}, and precise criteria for its presence, remain controversial.

\section{Methods}
\textit{\textbf{Simulation protocols.}}

Configurational ensembles for air-water and air-ice interfaces were sampled using molecular dynamics (MD) simulations performed with the LAMMPS\cite{lammps} MD package.
Forces were computed from the TIP4P/Ice model of water\cite{Abascal2005},
which was designed to accurately reproduce the thermodynamics of
phase coexistence between ice and liquid water\cite{Vega2005Phases,Abascal2005},
under periodic boundary conditions with Ewald summation of electrostatic interactions.
In all simulations, intramolecular bond distance constraints were imposed using the RATTLE algorithm\cite{Andersen1983} and constant temperature $T$ was maintained using a Nosé-Hoover thermostat with a damping frequency 200 fs$^{-1}$\cite{Hoover1985}.

The air-water simulation included $N=522$ molecules in a cell with dimensions $25\times25\times100\ \AAA^3$. The resulting liquid slab (initialized as a $25\times25\times25\ \AAA^3$ cube using the packmol\cite{packmol} package) spans the $x$ and $y$ dimensions of the simulation cell. 
This system was held at $T=298$ K and propagated with a time step of 2 fs, during both 4 ns of equilibration and a subsequent 5 ns production trajectory.

Air-ice simulations included $N=96$ molecules in a cell with dimensions $13.5212\times 15.6129 \times 14.72\ \AAA^3$. 
We constructed an initial ice slab by truncating an ideal, periodic ice crystal (generated according to the scheme of Hayward and Reimers\cite{Hayward1997}) so that the basal plane is exposed to vapor.
This system was advanced for 10 ns with a time step of 2 fs at a variety of temperatures. 
The system was made deliberately small in an effort to thoroughly equilibrate and broadly sample at low temperatures via parallel tempering. 
Simulation replicas were exchanged for temperatures at 10$\degree$C increments from 17 K to 287 K, with five additional simulations at 22 K, 32 K, 272 K, and 282 K. 
By exchanging replicas among 32 simulations, we explored a broad range of fluctuations even 135 K below the melting point.

As further assistance to sampling at low temperature, we hold the subsurface structure of ice fixed.
Over most of the temperature range that we consider, previous work suggests that L4 would be nearly static in the absence of constraints\cite{Conde2008, AlejandraSanchez2017, Kling2018}.
Accordingly, L4 is kept fixed throughout all ice simulations.

\vspace{0.1cm}
\noindent
\textit{\textbf{Instantaneous interface analysis.}}

For air-water interfaces, all depth-dependent properties are referenced to an instantaneous interface. For a given configuration the interface's height $z_{\rm inst}(x,y)$ is determined at a grid of lateral positions $(x,y)$ according to the coarse-graining scheme of Willard and Chandler\cite{Willard2010}.
A local, vapor-facing surface normal vector $\hat{\bm n}({\bm s})$ is then estimated at each surface grid point ${\bm s} = (x,y,z_{\rm inst}(x,y))$. 
The instantaneous depth of molecule $i$ is defined to be
\[
	d_i = (\bm{s} - \bm{r}_i) \cdot \hat{\bm n}(\bm{s}), \quad\quad\textrm{(air-water interface)}
\]
where $\bm{s}$ is the surface grid point closest to $\bm{r}_i$.
Throughout this work, the position ${\bm r}_i$ of a TIP4P/Ice molecule refers to the center of its excluded volume, i.e., the oxygen atom.
The air-water density profile we report is thus
$\rho_{\rm inst}(d) = (N/A)\langle \delta(d-d_i) \rangle$.

At temperatures of interest, undulations of the air-ice interface are quite limited. 
The static subsurface layer L4 in our simulations acts as an anchor that further removes uncertainty in the interface's location. 
In contrast to the soft liquid interface, the absolute vertical coordinate $z$ suffices as a detailed measure of depth in this case. 
We therefore define
\[
    d_i = z^{\rm L1A} - z_i + 0.8{\textrm \AA},
    \quad\quad\textrm{(air-ice interface)}
\]
where $z^{\rm L1A}$ denotes the crystalline height of the outermost half-bilayer of the basal plane. 
A shift of $0.8\ \AAA$ is introduced to facilitate comparison with the air-water interface.
The use of absolute height in the case of ice aids the detection and characterization of surface defects.

\vspace{0.1cm}
\noindent
\textit{\textbf{Orientational statistics.}}

We characterize the orientation of an interfacial water molecule by computing statistics of the angle $\theta_{\rm OH}$ between a hydroxyl bond vector ${\bm r}_{\rm OH}$ and the surface normal $\hat{\bm n}$. 
Specifically, we determine a depth-dependent probability distribution $P(\cos\theta_{\rm OH};d)$ of $\cos\theta_{\rm OH}$, and also a joint distribution $P(\cos\theta_{{\rm OH}_1},\cos\theta_{{\rm OH}_2};d)$ for the pair of OH bond vectors that together define a water molecule's orientation up to an angle of azimuthal symmetry.
The singlet distribution is uniform in an isotropic bulk liquid (or vapor).
The joint distribution, however, is nonuniform even in an isotropic environment (an exact calculation for the bulk case is given in SI). We follow previous work in referencing $P(\cos\theta_{{\rm OH}_1},\cos\theta_{{\rm OH}_2};d)$ to its bulk form.

\vspace{0.1cm}
\noindent
\textit{\textbf{Lateral correlation of ice-like motifs.}}

In following sections we will show that a local structural motif characteristic of the air-ice interface is recapitulated in detail at the air-water interface. 
For the idealized crystal surface, this arrangement comprising a few molecular centers is repeated along lattice vectors parallel to the surface, in perfect registry, forming the outermost layers of ice's basal face.
At a liquid surface, which is azimuthally symmetric, an average long-range alignment of local motifs is prohibited.
Resemblance between the two surfaces is thus necessarily limited to a microscopic scale.
We assess this length scale through a correlation function $S(r)$ that quantifies motifs' alignment as a function of lateral distance $r$.

The repeated structural element we consider centers on a molecule in the outermost bilayer (L1), which ideally forms hydrogen bonds with three other molecules in L1.
To characterize the relative alignment of two such motifs, 
we first project the position of each L1 molecule (those with $0<d<1.3$ \AAA) onto the local instantaneous interface,
\[
\tilde{\bf r} = {\bf r} - ({\bf r} \cdot \hat{\bf n})\hat{\bf n}.
\]
For the idealized crystal, these projected positions lie at the vertices of a honeycomb lattice in two dimensions.
For the liquid surface, we anticipate honeycomb-like domains, limited in size by orientational decoherence and the presence of structural defects. 

Bond orientation parameters,
\[
	q_m &= \frac{1}{N_{\rm b}(m)} \sum_{j=1}^{N_{\rm b}(m)} e^{3i \theta_{mj}},
\label{equ:steinhardt}
\]
akin to those of Steinhardt and Nelson\cite{Steinhardt1983} allow a simple statistical analysis of these domains. 
The sum in Eq.~\ref{equ:steinhardt} runs over the $N_{\rm b}(m)$ surface neighbors of a surface molecule $m$.
(Neighbors are identified by intermolecular distance, $|\bm r_m - \bm r_j|<3$ \AA, a range encompassing all strong hydrogen bonds.)
$\theta_{mj}$
is the angle between the bond vector $\tilde{\bf r}_{mj} = \tilde{\bf r}_m  -\tilde{\bf r}_j$ and an arbitrary
lateral axis in the laboratory frame. For a pair of surface molecules $m$ and $k$, the product $q_m q_k^*$
indicates the alignment of the respective local coordination environments. 
A perfect honeycomb
gives $q_m q_k^* = \pm 1$, depending on the vertices occupied by $j$ and $m$. Our order parameter for the
alignment of ice-like surface motifs is a conditional average of this product,
\[
S(r) = \frac{\braket{q_m q_k^* \delta(r-|{\bf r}_{mk}|)} } {\braket{\delta(r-|{\bf r}_{mk}|)}}, \label{Eq:S_of_r}
\] 
for surface molecules separated by a distance $r$.
Uncorrelated random bond orientations give $S(r)=0$, while the ideal basal face of ice gives
a series of values $S(r)=\pm 1$ at discrete distances corresponding to vertex pairs in the projected honeycomb lattice.

\section{Results and Discussion}

To establish connections between the surfaces of water's liquid and solid phases, we will compare three distinct systems. 
One is an ideal crystal of ice Ih terminated at a specific lattice plane.
In the included figures, the precisely defined features of an ideal crystal surface will be indicated by black lines and points.

The second system we will consider is a finite-temperature crystal in coexistence with vapor. 
In figures, we show simulation results for $T=137$ K,
the lowest temperature that is frequently accessed in our parallel tempering by thoroughly exchanging replicas.
This system is thus well equilibrated (within the constraint of fixed subsurface molecules in L4), but sufficiently cold that the ice surface is highly organized. 
Substantial deviations from the ideal crystal structure are in most cases well localized in space (aside from occasional transitions of the entire surface to a cubic ice arrangement). 
By most standards a quasi-liquid layer, whose fluidity and structural variability are reminiscent of ambient liquid water, is absent under these conditions. 
Simulation results for this finite (but low) temperature system are presented in Figs. 2, 5, and 6 as blue lines.

Our final system is the much-studied air-water interface, as represented by the TIP4P/Ice model at 298 K.
Simulation results for the liquid surface are shown in Figs. 2, 5, and 6 as red lines.

We characterize each of these three systems in several ways, emphasizing different aspects of microscopic organization at aqueous surfaces.
We first examine density profiles that highlight discrete molecular layering, made evident at the liquid surface by referring depth to an instantaneous interface.
We then compare orientational structure of the revealed layers, focusing specifically on statistics of molecular dipoles and 
of a water molecule's two OH bond vectors. 
Finally, we quantify the degree of lateral alignment among molecular coordination environments in each system's outermost layer.
These comparisons paint a consistent picture: The ice surface's quasi-liquid layer, a semi-structured region
in which crystalline organization is locally evident but globally indistinct, is an apt description of the air-water interface as well.

\vspace{0.1cm}
\noindent
\textit{\textbf{Depth profiles of molecular density.}}

\begin{figure}[!t]
	\includegraphics[width=\linewidth]{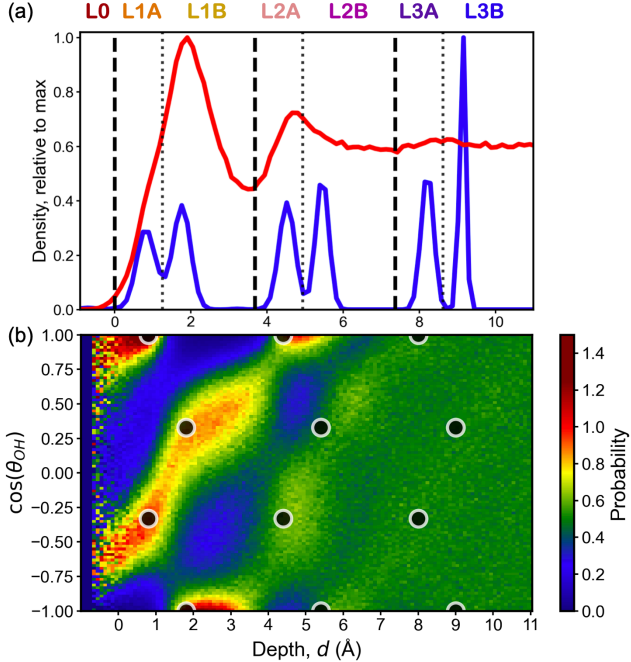}
	\caption{ 
	Layering of molecular density and orientational patterns at the air-water and air-ice interfaces. (a) Density profiles
	for the two surfaces, scaled by their maximum values over the range shown. In the liquid-vapor case (red), depth is measured relative to the instantaneous interface. Ice results are shown for $T=137$ K (blue). Black lines indicate layer and sublayer divisions in an ideal ice lattice. (b) Probability distributions of molecular orientation at the liquid surface (colors). $P(\cos\theta_{\rm OH};d)$ is normalized separately at each depth $d$. Black circles show expectations from an ideal lattice.
    }
	\label{fig:density-ice-water}
\end{figure}

An ideal crystal truncated at a lattice plane exhibits distinct molecular layers parallel to the surface, separated by distances that are dictated  by the lattice structure.
In the case of ice's basal face, these strata group naturally into a series of evenly spaced bilayers
(L1, L2, L3, $\ldots$), as described above and illustrated in Fig.~\ref{fig:two-layer-schematic}.
The 0.96 \AAA\ spacing within one of these bilayers (i.e., the vertical distance between L$X$A and L$X$B peaks) is several
times smaller than the 3.68 \AAA\ spacing between adjacent bilayers (i.e., the vertical distance from L1A to L2A).
Although ice is built from the same tetrahedral hydrogen bonding motif as the liquid, neither of these separation distances 
correspond closely to lengths that are familiar from standard measures of liquid structure, e.g., peaks in the oxygen-oxygen radial distribution function $g_{\rm OO}(r)$ at 2.81 \AAA, 4.46 \AAA, $\ldots$\cite{Piaggi2019}.
Our simulations of the ice surface at low but finite temperature yield density profiles (Fig.~\ref{fig:density-ice-water}, blue) whose peak positions follow that of the ideal crystal lattice. 
At 137 K bilayer substructure is clearly evident, although
broadened L1A and L1B peaks overlap significantly.

The density of liquid water at its interface with air is also structured, as Willard and Chandler strikingly
demonstrated\cite{Willard2010}. 
Fig.~\ref{fig:density-ice-water} shows the results of such an analysis in red for the TIP4P/Ice model at liquid-vapor coexistence.
This depth profile closely matches those calculated previously for other point-charge models of water\cite{Willard2010,Sega2015,Serva2018} and agrees qualitatively with profiles computed by ab initio molecular dynamics\cite{Kessler2015}.
Two pronounced peaks appear, at $d\approx 1.9$ \AAA\ and $d\approx 4.7$ \AAA.
The first marks a well-defined outermost molecular layer, and has a shape that loosely suggests structure within the layer.
Specifically, the increase from gas to liquid density is not uniformly steep and includes multiple inflection points. 
In addition, and in sharp contrast to $g_{\rm OO}(r)$, the first density peak is at least as broad as the second. 
A subtle third peak can just be resolved near $d\approx 8.5$ \AA.

Such characterizations of the air-water interface have inspired several detailed layer-by-layer analyses\cite{Partay2008,Sega2015,Kessler2015,Zhou2017,Pezzotti2017, Pezzotti2018,Serva2018,Shin-Willard}.
The precise boundaries of each layer differ among studies, but some important general conclusions are consistently reached. 
Kessler et al.\cite{Kessler2015} separated the liquid water density profile into layers 3 \AAA\ in width (denoted L0, L1, and L2), and designated the upper 1.2 \AAA\ of L1 as L1$^\parallel$ due to a unique internal structure that we will discuss later.
They further observed that orientations, hydrogen bonding, and residence time differed measurably in each layer\cite{Kessler2015}.
Gaigeot and coworkers extended the layer-by-layer analysis\cite{Pezzotti2017,Pezzotti2018,Serva2018} by noting substantial intra-layer connectivity, especially within L1$^\parallel$.

The similarity of the liquid and icy density profiles in Fig.~\ref{fig:density-ice-water} suggests that the layering at the liquid's surface may be similar in nature to the crystalline structure at ice's basal face.
In particular, peaks of the liquid profile, designated in previous work as L1, L2, $\ldots$, align well with the ice bilayers we have denoted L1, L2, $\ldots$.
We therefore define the boundaries of liquid surface layers with reference to the corresponding layers of ice, which are spaced by 3.68 \AAA.
Setting the boundary between L1 and L2 at $d=3.68$ \AAA\ (very near the first local minimum of $\rho_{\rm inst}(d)$), we arrive at the layer divisions indicated in Fig.~\ref{fig:density-ice-water}a by dashed lines.

Substructure within the ice bilayers can further help to explain peculiarities of the air-water density profile.
We associate the shoulder of the liquid's first peak with L1A molecules of ice's basal face, and the main peak with L1B.
Setting the boundary between L1A and L1B at $d=1.3$ \AAA\ (near an inflection point where the peak shoulder ends), we obtain the sublayer divisions indicated in Fig.~\ref{fig:density-ice-water}a by dotted lines.
For the ideal ice lattice, populations of L1A and L1B are equal.
In the analogy we are proposing, substantial density has shifted from the liquid's L1A into L1B, giving a twofold difference in sublayer-averaged densities, $\rho_{\rm inst}^{\rm L1A} / \rho_{\rm inst}^{\rm L2A} = 0.53$.
(Table~SI.2).
The net density of L1 in our definition is slightly lower than that of other layers, $\rho_{\rm inst}^{\rm L1} / \rho_{\rm inst}^{\rm L2} = 0.92$.

Relative peak widths in the liquid density profile can be rationalized within the ice analogy.
The distinctness of molecular layers rapidly attenuates moving toward the translationally symmetric bulk liquid.
Density peaks are expected to broaden as a result, just as successive peaks in $g_{\rm OO}(r)$ broaden with increasing distance.
Bilayer substructure, however, enhances the peak widths of layers that are most ordered.
L1A and L1B are separated in depth at the liquid surface, even if their density contributions strongly overlap.
As crystalline order attenuates moving toward the bulk liquid, bilayers' internal separation weakens.
Resultant merging of A and B sublayers acts to reduce the width of a layer's net density peak. 
The comparably broad peaks observed for L1 and L2 in simulations could be viewed as a consequence of these countervailing trends.

\begin{figure*}[t]
	\centering
	\includegraphics[width=\linewidth]{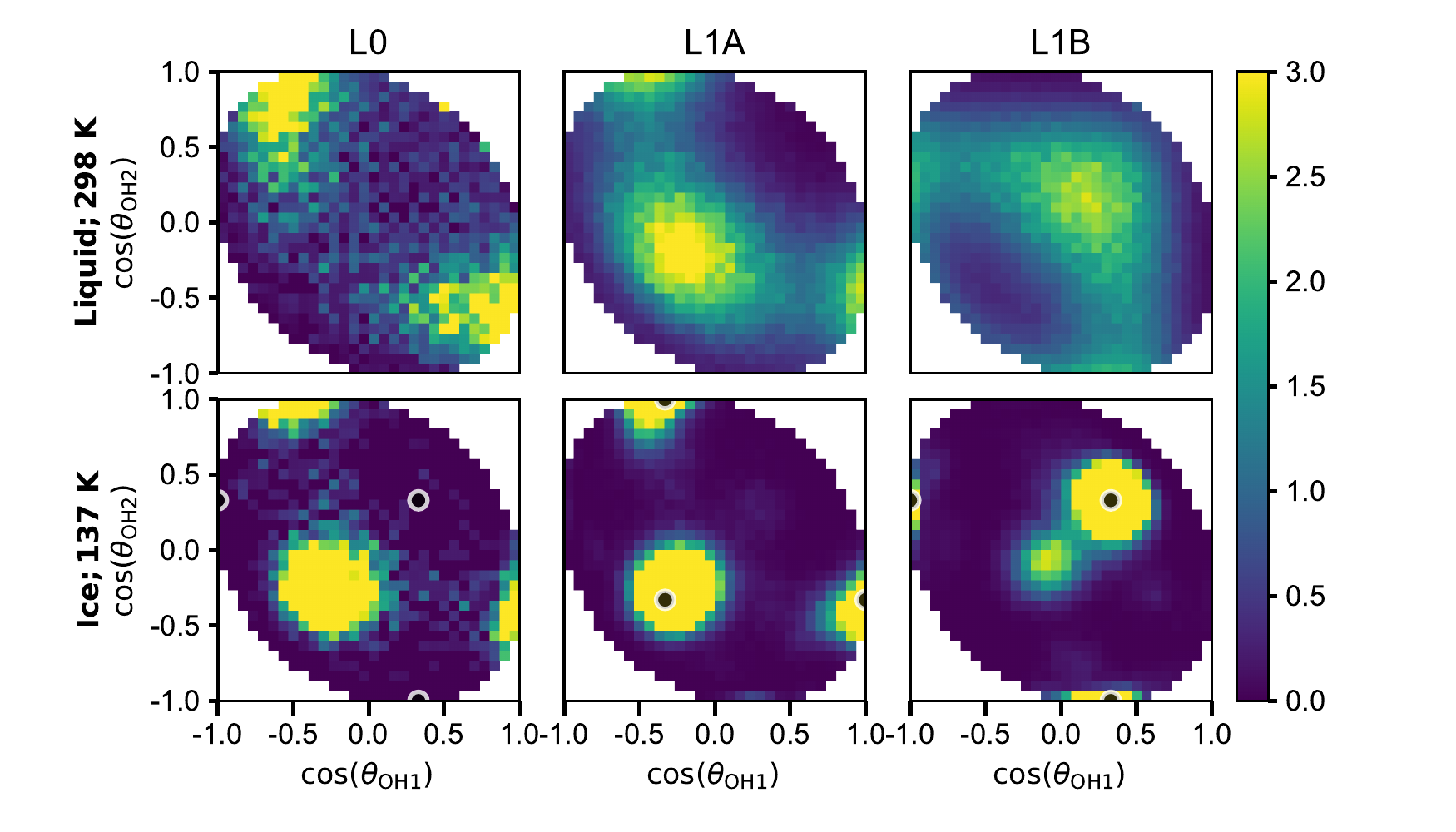}
	\caption{
	Joint probability distributions $P(\cos\theta_{{\rm OH}_1},\cos\theta_{{\rm OH}_2};d)$ for the orientations of a water molecule's two OH bond vectors, computed for liquid-vapor (top) and ice-vapor (bottom) interfaces. Each plot shows results aggregated for a specific sublayer (L0, L1A, or L1B) and scaled by the probability distribution for an isotropic bulk environment. Black circles show expectations for these sublayers from an ideal lattice, including L0 as a continuation of the periodic L$X$A / L$X$B pattern.
	}
	\label{fig:coscos-summary}
\end{figure*}

Other faces of ice Ih also feature characteristic layering, but their density profiles do not align with the liquid surface as well as the basal face. The density profile of the primary prismatic and secondary prismatic faces of ice is included in %Figs.~SI.\ref{fig:density-PP} and SI.\ref{fig:density-SP}. 
Figs.~SI.2 and SI.3. 
The primary prismatic face, whose boat hexagons are closely related to chair hexagons at the basal face, exhibits similar layering but at slightly wider intervals than the basal plane. 
The secondary prismatic face is an especially poor match to the layering of the liquid surface.
The orientational structures at the other ice faces also do not match 
the liquid as closely as the basal face (compare, e.g., Fig.~\ref{fig:coscos-summary} to
% Fig.~SI.\ref{fig:prismatics-orientations}).
Fig.~SI.5).

\vspace{0.1cm}
\noindent
\textit{\textbf{Orientational statistics of water's outermost layers.}}

At temperatures of interest ice Ih is proton disordered, so that the directions along which a particular water molecule donates hydrogen bonds are not completely determined. These directions are nonetheless highly constrained by the ordered molecular lattice, allowing only four possible OH bond vector orientations for each molecular lattice site.
At the basal surface of ice, allowed orientations reflect the puckered hexagonal motif that tessellates to form each bilayer of ice. 
For the ideal crystal surface, water molecules in L1A sacrifice one potential hydrogen bonding site directly to vapor, while forming hydrogen bonds with their three closest neighbors, all in L1B. The angle $\theta_{\rm OH}^{\rm L1A}$ thus has two possible values, satisfying $\cos \theta_{\rm OH}^{\rm L1A} \approx +1$ and $\cos \theta_{\rm OH}^{\rm L1A} \approx -1/3$, for outward- and inward-pointing OH groups, respectively. Likewise, every water molecule in L1B forms three hydrogen bonds with neighbors in L1A, $\cos \theta_{\rm OH}^{\rm L1B} \approx +1/3$, and forms a fourth hydrogen bond pointing directly away from the surface, $\cos \theta_{\rm OH}^{\rm L1B} \approx -1$, that connects to the subsequent layer, L2A. The oscillatory upward-reaching (L$X$A) and downward-reaching (L$X$B) structure repeats in each ice layer.
This pattern is indicated in Fig. \ref{fig:density-ice-water}b by black circles at allowed angles for each molecular depth in the ideal lattice.

Molecules at the air-water interface show a very similar pattern of orientational preferences. 
The singlet distribution $P(\cos\theta_{\rm OH};d)$, plotted in Fig.~\ref{fig:density-ice-water}b, exhibits a series of peaks that align well with the collection of allowed hydrogen bond orientations at ice's basal face.
Relative to ice, peaks of $P(\cos\theta_{\rm OH};d)$ are broad at the liquid's surface, with FWHM $\approx30\degree$ throughout L1 and broader still for L2.
These peaks are most pronounced at depths shifted from ice-based expectations by about 0.5 \AAA\ in L1B and L2.
An ice-like progression of features near $\cos\theta_{\rm OH}=\pm 1$ and $\pm 1/3$ is nonetheless unmistakable. 

Some of these orientational similarities have been discussed in earlier work.
In particular, the surface-exposed ``dangling'' OH ($\cos\theta_{\rm OH}= +1$ in L1A) has been demonstrated spectroscopically\cite{Du1993,Du1994,Fan2009}, and the corresponding outward-facing OH of the basal plane of ice was noted by Du et al.\cite{Du1994}.
Fan et al.\cite{Fan2009} further observed in MD simulation that a second layer of water tends to point one OH toward bulk ($\cos\theta_{\rm OH}= -1$ in L1B). 
By accounting for surface shape fluctuations, the comparison in Fig.~\ref{fig:density-ice-water}b is much more precise than in previous studies.

The more detailed joint distribution $P(\cos\theta_{{\rm OH}_1},\cos\theta_{{\rm OH}_2};d)$ reinforces this close similarity in orientational structure. 
For the ideal crystal, each of the sublayers L1A, L1B, L2A, \ldots allows three possible ordered pairs $(\cos\theta_{{\rm OH}_1},\cos\theta_{{\rm OH}_2})$, namely $(1,-1/3)$, $(-1/3,1)$, and $(-1/3,-1/3)$ in L$X$A, and $(-1,+1/3)$, $(+1/3,-1)$, and $(+1/3,+1/3)$ in L$X$B, indicated by black circles in Fig.~\ref{fig:coscos-summary}. 
Simulations of ice at 
%the finite temperature 
137 K yield peaks near these expected values for L1A and L1B, broadened by thermal fluctuations. 
For L1A at the liquid surface, probability is highest in the same intervals of $(\cos\theta_{{\rm OH}_1},\cos\theta_{{\rm OH}_2})$ as in ice, and likewise for L1B. 
This agreement is demonstrated in Fig.~\ref{fig:coscos-summary} by plotting sublayer-resolved histograms, which integrate $\rho_{\rm inst}(d) P(\cos\theta_{{\rm OH}_1},\cos\theta_{{\rm OH}_2};d)$ over the range of $d$ defining a given sublayer. The peaks of these histograms align well for all three of our systems in L1A, and also in L1B. Peaks are unsurprisingly broader and less pronounced in the liquid case.
\pagebreak[4]

\vspace{0.1cm}
\noindent
\textit{\textbf{Ice-defect features at the liquid surface.}}

\begin{figure}[!t]
	\centering
	\includegraphics[width=\linewidth]{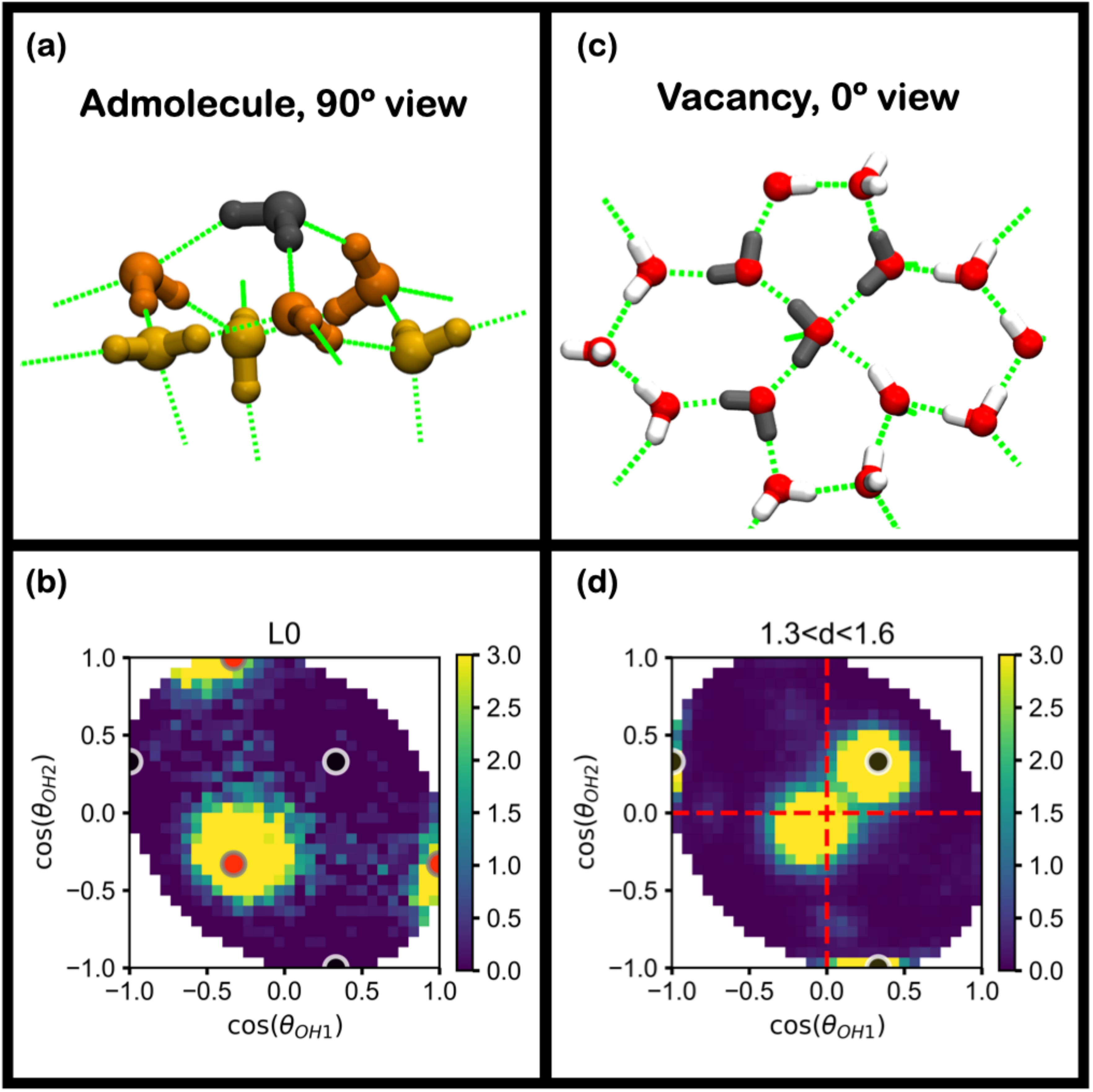}
	\caption{
	Discrete defects in ice surface structure that strongly influence statistics of molecular orientation at $T=137$ K.
	Hydrogen bonds are indicated in schematics (a) and (c) by green dotted lines. 
	Joint probability distributions $P(\cos\theta_{{\rm OH}_1},\cos\theta_{{\rm OH}_2};d)$ in (b) and (d) are averaged
	over the depth range indicated and scaled by the bulk result, and black circles indicate
	expectations from the ideal ice lattice.
	(a) Surface self-interstitial defect viewed from the side (the surface normal $\hat{\bf n}$ points upward).
	Three members of L1A (orange) each form one hydrogen bond with the admolecule (black) in L0, which sits atop the L1 hexagon as the apex of a pyramid. (b) Distribution of molecular orientation in the L0 layer of ice.  Red circles show orientational preferences of the admolecule defect. (c) Top-down view of a vacancy defect (the surface normal $\hat{\bf n}$ points out of the page) at the basal surface of ice. Only molecules in L1 are shown. In this snapshot of a highly variable defect structure, one water (center) forms five strained hydrogen bonds, which require the nearby molecules (gray) to lie nearly parallel to the interface and to reside at a depth between the ideal L1A and L1B sublayers. (d) Distribution of molecular orientation in a narrow range of depth near the L1A/L1B boundary.
}
	\label{fig:defects-4panel}
\end{figure}

Joint angle distributions for finite-temperature ice differ from ideal lattice expectations in several ways.
Because the ice surface at 137 K remains highly ordered overall, we expect that these differences are associated with discrete defects in crystal structure.
Simulations of rigid point-charge water models have revealed a handful of such structural defects on the pristine ice basal surface\cite{Buch2008,Bishop2009,Watkins2011,Pedersen2015,Slater2019}, whose lifetimes and energies have been quantified\cite{Watkins2011,Pedersen2015,Slater2019}. 
These defect states preserve or nearly preserve the total number of hydrogen bonds, at the cost of distortions to the idealized tetrahedral bond geometry of ice. 
With a low energetic cost, they serve as the primary source of discrete fluctuations at the air-ice interface for low but finite temperatures.
Two of these discrete defects, depicted in Fig.~\ref{fig:defects-4panel}, are particularly helpful for understanding the orientational statistics we have computed.

In the first defect, a surface self-interstitial, a water molecule sits above the outermost bilayer, forming three hydrogen bonds with L1A molecules. This admolecule does not simply continue the periodic structure of the layers beneath it; in such a continuation it would form only one hydrogen bond with L1A, incurring a large energetic penalty.
Instead, the admolecule sacrifices a single bond and, as a result, is oriented much as in L$X$A. Red circles in 
Fig.~\ref{fig:defects-4panel}b show these expectations for an admolecule defect, which align very closely with our simulation results for L0 in finite temperature ice.
As a more subtle signature of this defect, the L1A molecules that bond with the admolecule must shift slightly out of their regular tetrahedral orientation, consistent with faint features near $\pm$(0.75, -0.4) in the ice L1A panel of Fig.~\ref{fig:coscos-summary}.

Our results for L0 at the liquid surface show a remarkably similar pattern.
In the liquid case, L0 samples are rare almost by definition: The instantaneous interface would be deformed upwards by an admolecule, which could well be assigned to L1 as a result.
A population of strongly protruding molecules is nonetheless present, and its orientational preferences are most reminiscent of L$X$A.
By comparison with finite-temperature ice, a peak at $(-1/3,-1/3)$ is missing in the joint distribution for liquid L0.
We attribute this absence to the awkwardness of detecting admolecules with a Willard-Chandler interface, which could assign the corresponding population to L1A instead.

The second discrete defect we consider is a vacancy in L1. 
The outright removal of an L1A molecule from the ideal crystal surface severs three hydrogen bonds.
To offset this loss, surrounding L1A molecules rearrange to form a number of new bonds that are strained. 
This rearrangement is not precisely defined, introducing a variable mix of polygons of degree 4 or higher\cite{Bishop2009}.
Fig.~\ref{fig:defects-4panel}c shows an example vacancy structure, in which one water forms five strained hydrogen bonds. 
This molecule and several of its neighbors lie nearly parallel to the surface and sit at a depth near the L1A/L1B boundary.

The L1B joint angle distribution for finite-temperature ice exhibits a peak near $(0,0)$,
which we associate with the vacancy defect. In this orientation, a water molecule's nuclei all lie in a plane parallel to the interface.
The joint histogram in Fig. \ref{fig:defects-4panel}d, which is limited to depths near the L1A/L1B boundary, confirms that parallel orientation is strongest in this inter-sublayer region of the ice surface.

A parallel surface region (L1$^\parallel$) has already been reported for the liquid surface\cite{Willard2010,Ishiyama2012,Kessler2015,Pezzotti2017,Pezzotti2018,Serva2018}, with many similarities to the defective ice structures we have described.
As in ice, the sublayer with enhanced parallel character is situated near the L1A/L1B boundary. 
Like the vacancy defect structure, the parallel sublayer exhibits many hydrogen bonds between molecules in the parallel region\cite{Kessler2015,Pezzotti2018,Serva2018}.
The joint angle distributions we have calculated make clear that the parallel region in ice is a small sub-population of L1, made prominent only by limiting attention to the narrow strip of depths between L1A and L1B.
Assessing the population of the liquid's parallel region, however, is made difficult by a lack of sharp features in distributions of depth or orientation.
Indeed, previous definitions of this region vary significantly, as do the resulting counts of molecules it includes.
In SI we examine several classification criteria, their physical basis, and the molecular populations they imply.
This analysis classifies roughly 10-30\% of molecules in liquid L1 as parallel in character.

\vspace{0.1cm}
\noindent
\textit{\textbf{Sub-bilayer surface structure.}}

\begin{figure}[!t]
	\includegraphics[width=\linewidth]{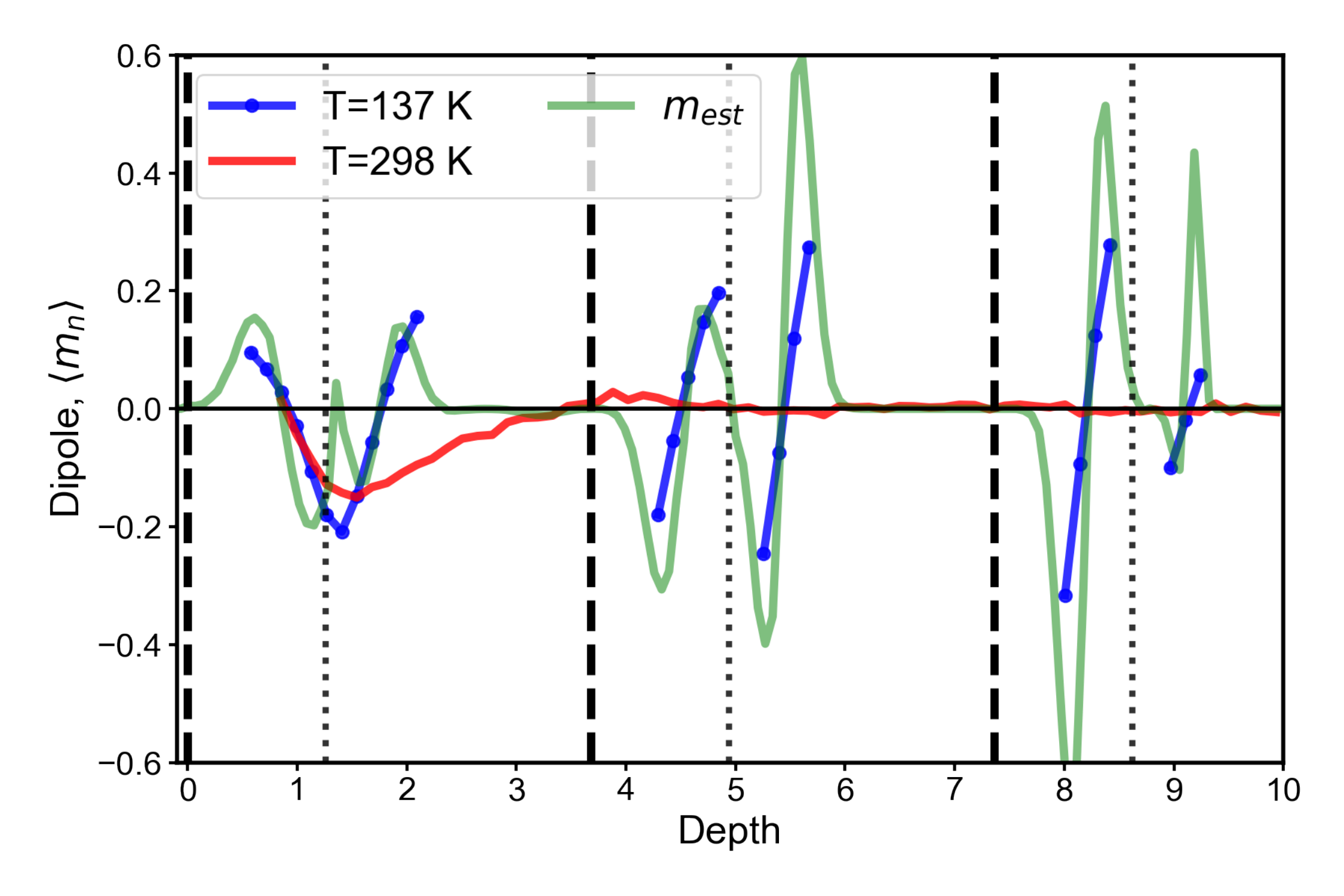}
	\caption{
	Average $z$-component of the molecular dipole moment, $\braket{m_{\rm n}(d)}$, computed as a function of depth for the ice surface (137 K, blue) and liquid water surface (298 K, red). Dashed and dotted black lines indicate ice layer boundaries
	just as in Fig.~\ref{fig:density-ice-water}. Ice results are shown only at depths where density is sufficient to obtain reliable averages. Systematic sub-bilayer dipole structure is reproduced qualitatively by $m_{\rm est}$ (green), the result of a simple model (see SI) that distinguishes molecules by the layers in which their hydrogen bonding partners reside. 
	}
	\label{fig:dipole-water-iceT-135}
\end{figure}

The dipole ${\bf m}$ of a water molecule -- a linear combination of its two OH vectors -- obeys statistics that are completely determined by the joint distribution $P(\cos\theta_{{\rm OH}_1},\cos\theta_{{\rm OH}_2};d)$. 
The depth-resolved average dipole $\braket{m_{\rm n}(d)} = \braket{{\bf m}\cdot \hat{\bf n}({\bf s})}$ at a surface nonetheless reports on hydrogen bond network properties that are only subtly encoded in this distribution. Simulation results for $\braket{m_{\rm n}(d)}$ reveal yet another similarity between liquid water and ice surfaces. 
Any model of tetrahedral bonding units whose donating and accepting sites are statistically equivalent gives $\braket{m_{\rm n}(d)}=0$ as a requirement of symmetry.
For models like TIP4P/Ice, the average molecular dipole therefore reports on the statistical differences between donated and accepted hydrogen bonds. 
Fig.~\ref{fig:dipole-water-iceT-135} shows the dipole profile computed for the liquid's surface (red), which closely resembles previous results for similar models\cite{Shin-Willard}. 
We also show the dipole profile for finite-temperature ice (blue), which recapitulates the most prominent features of the liquid result. Specifically, in both cases $\braket{m_{\rm n}(d)}$ crosses zero in L1A and is minimum near the L1A/L1B boundary. 
For ice, patterned structure in $\braket{m_{\rm n}(d)}$ continues many layers from the interface, while for the liquid $\braket{m_{\rm n}(d)}$ is distinguishable from zero only in L1 and L2 ($d < 7$ \AAA). 

In the SI, we argue that the ice result indicates distortions in each sublayer that systematically displace upwards-pointing OH groups in L$X$A towards the bulk, and downwards-pointing OH groups in L$X$B towards the surface, with the exception of L1A where the displacement direction is reversed. 
For the liquid surface, our ice analogy thus suggests that the dipole profile arises from subtle but systematic displacements of water molecules according to the direction of their hydrogen bonds and the sublayers they occupy.
We also propose a minimal model, $m_{\rm est}(d) = \sum_i \bar{m}_i^{\alpha(d)} \rho_i(d)$ (see SI 
%Eq.~\ref{equ:mest}
Eq.~SI.12), that captures the essential surface dipole structure as simple deviations from L$X$A and L$X$B positions.

\vspace{0.1cm}
\noindent
\textit{\textbf{Lateral decoherence of ice-like structure.}}

Structure parallel to the interface is slightly more challenging to characterize than layering in the perpendicular direction of broken symmetry. 
The correlation function $S(r)$ in Eq.~\ref{Eq:S_of_r} is designed for this purpose, quantifying the lateral alignment of local structural motifs that in ice are coherently arranged over macroscopic distances.

\begin{figure*}[t]
	\centering
	\includegraphics[width=\linewidth]{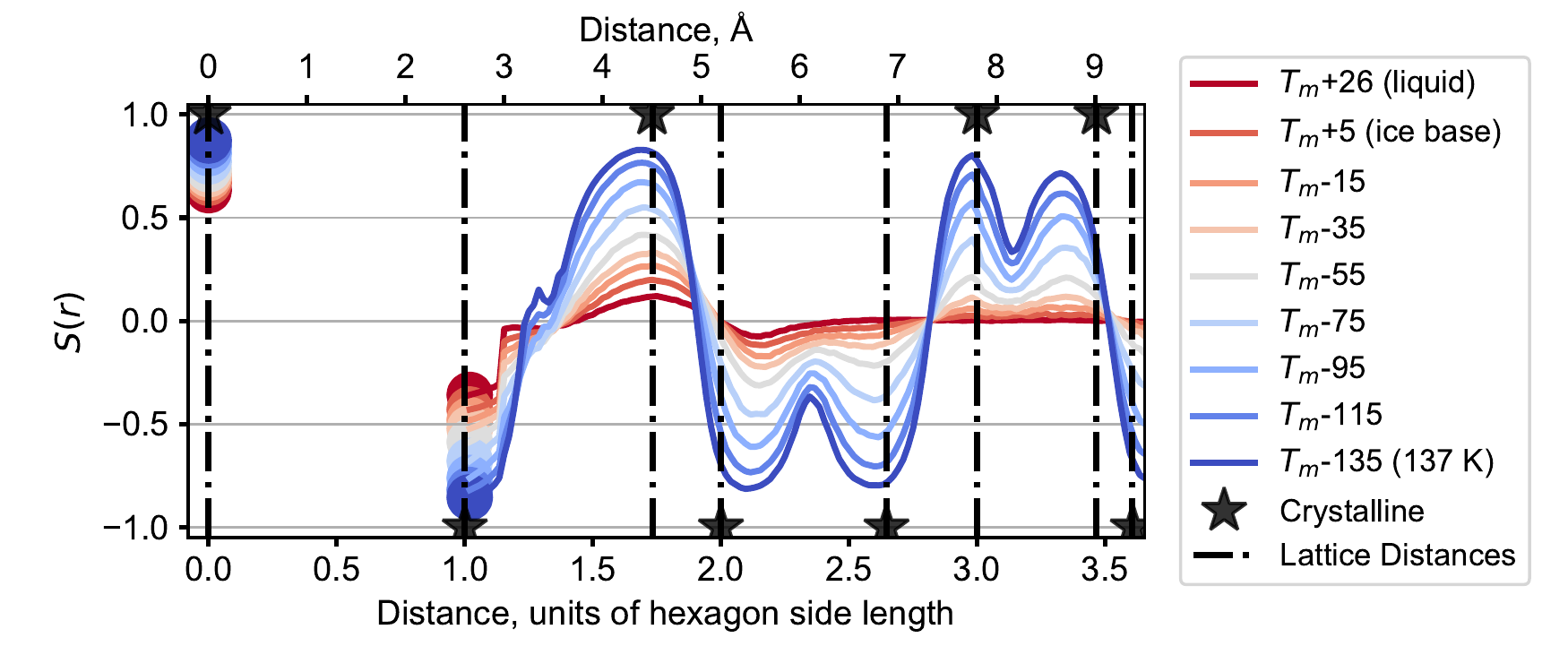}
	\caption{
	Lateral alignment of hydrogen bonding geometries in the range of depth corresponding to the L1 layer of ice. Results for the bond order parameter correlation function $S(r)$ are shown for ice at a range of temperatures spanned by parallel tempering simulations with the L4 subsurface layer held fixed (including a temperature just above $T_{\rm m}$, at which the system would melt without the ice base constraint). The liquid result at 298 K was obtained without subsurface constraints.
	Stars indicate ideal lattice expectations at vertex-vertex distances 
	$r = \ell [(\sqrt{3} i_x/2 )^2 + ( i_y/2 )^2]^{1/2}$
	of the hexagonal lattice that results from projecting the L1 layer onto the plane of the interface. $i_x$ and $i_y$ are integers whose sum must be even, and $\ell = 2.6$ \AA\ is the side length of projected hexagons.
	Sharp changes at $r=1.15 \ell$ coincide with the cutoff for identifying bonding partners that contribute to each molecule's bond orientation parameter $q_m$.
	}
	\label{fig:registers-byz-selectedt}
\end{figure*}

For two lattice sites in the L1 layer of an ideal basal face, separated by a distance $r$, the correlator $S_{\rm ideal}(r)$ has unit magnitude and a sign determined by the sublayers they inhabit: $S_{\rm ideal}(r)=+1$ for two sites in L1A (also for two sites in L1B), while $S_{\rm ideal}(r)=-1$ for sites in different sublayers. 
The resulting pattern of $\pm 1$ values is indicated in Fig. \ref{fig:registers-byz-selectedt} by stars. At 137 K this pattern is closely followed. The overall scale of $S(r)$ is slightly diminished, even at $r=0$, because local hydrogen bonding environments are not perfectly equiangular, but strong positive and negative features appear very near those of the ideal lattice. Excursions of molecules away from their ideal lattice sites broaden these features. Such excursions also generate destructive interference where $S_{\rm ideal}(r)=-S_{\rm ideal}(r')$ at nearby distances $r$ and $r'$, so that peaks and troughs in $S(r)$ are shifted slightly away from their ideal locations.

At the liquid surface, $S(r)$ follows the very same pattern of positive and negative features as for finite-temperature ice, but its magnitude decays with $r$ over a scale of $\approx 7$ \AAA\ . This resemblance demonstrates that ice-like structure at the air-water interface is not limited to vertical layering of density and orientational preferences of individual molecules. Lateral honeycomb patterning akin to ice's basal face is clearly evident in $S(r)$, with a coherence length of 2-3 molecular diameters. This length scale is nearly identical to the depth of coherent layering demonstrated by $\rho_{\rm inst}(d)$ and $P(\cos\theta_{\rm OH};d)$, which decay completely within 7-8 \AAA\ of the liquid's outermost layer.

Fig. \ref{fig:registers-byz-selectedt} shows results as well for $S(r)$ at several temperatures spanning the range from 137 K to ambient temperature. The smooth progression of $S(r)$ with increasing $T$ casts the liquid surface as the endpoint of a gradual disordering of ice that begins already at 137 K. At such low temperature this decorrelation can only be driven by small-amplitude vibrations and sparse discrete defects. Sufficiently accumulated at higher $T$, individual defects are hardly distinct, but Fig. \ref{fig:registers-byz-selectedt} suggests that their accumulation indeed underlies the surface structure of ice near melting. The resemblance between $S(r)$ for ice at $T\approx T_{\rm m}$ and for liquid at 298 K is striking, and would likely be stronger still if constraints on subsurface layers were relaxed in our simulations. Liquid and high-$T$ solid surfaces thus appear to share the semi-structured exterior often described as a quasi-liquid layer on ice. From the perspective advanced in this paper, it could equally well be described as a quasi-ice layer on the surface of liquid water.

\section{Conclusion}
We have presented results of computer simulations that strongly support an analogy between the air-ice and air-water interfaces.
The analogy is not meant to suggest extended periodic structure at the liquid's surface, but instead to highlight the presence of nanometer-scale domains in which molecular layering, orientation, and hydrogen bond arrangements mimic those at the basal face of ice.

This conclusion echoes previous work that pointed less strongly to similar conceptions of liquid water's surface, but with important amendments. 
Lee et al.\cite{Lee1984} posited that external forces which drive close molecular packing are an essential ingredient for the formation of ice-like layering at a hydrophobic substrate.
We have shown that the necessary driving forces are in fact intrinsic to aqueous interfaces, even without a hard wall to pack against. 
The signatures and consequences of these forces, however, are pronounced only in an instantaneous interface analysis that accounts for shape fluctuations of soft interfaces. 
Fan et al.\cite{Fan2009} drew a conclusion similar to ours for molecular organization perpendicular to the air-water interface, but viewed such organization to be absent in parallel directions.
We have shown that correlation lengths for ice-like structure are in fact nearly equal in perpendicular and parallel directions.
Lateral organization, however, is evident only in those observables that break symmetry appropriately, such as the two-point correlation function $S(r)$ for projected bond order parameters.

The ice-like structure we have reported at the surface of liquid water could be viewed as a phenomenological mirror image of the quasi-liquid layer at the surface of ice.
For ice, the interface with vapor necessarily severs many hydrogen bonds within the ideal lattice structure, allowing a set of low-energy crystal defects to be populated even at fairly low temperature.
From this perspective, the quasi-liquid layer constitutes a buildup of low-energy defects 
that become increasingly common at modest temperatures\cite{Bishop2009}, consistent with changes in
VSFS peak amplitudes over a broad range of temperature\cite{Shen2006,Smit2017B,Tang2018,Tang2020}.
In the opposite way, a vapor interface with liquid water offers few ways to satisfy a significant fraction of hydrogen bonds. 
The limited collection of low-energy liquid surface configurations results in a kind of quasi-ice layer at the surface of water.

The analogy between ice and liquid water presented here neglects known\cite{Bjerrum1952,Watkins2010} non-classical defects (i.e., Bjerrum defects) and is based on simulations of fully classical, non-reactive, rigid point-charge models of water. 
Neglected effects of electronic polarization, molecular distortions, and proton disorder certainly impact surface structure and thermodynamics.
Nevertheless, rigid water models reproduce phase behavior\cite{Vega2005Phases} and liquid spectra\cite{Schmidt2007}, can match ice geometry, and provide reasonable estimates of surface tension\cite{Vega2007}.  
Ab initio simulations of the water surface measure a surface structure that is similar to simpler point-charge models\cite{Leontyev2011,Kessler2015,Pezzotti2017,Pezzotti2018,Besford2018,Wohlfahrt2020}. 
Given the strong connections we have demonstrated between the air-ice and air-liquid surfaces in a rigid point-charge model,
we expect that detailed electronic effects will be secondary to the pronounced consequences of molecular geometry that are common to any reasonable microscopic description of water.

\section{Acknowledgments}
This work was supported by the U.S. Department
of Energy, Office of Basic Energy Sciences, through the Chemical
Sciences Division (CSD) of Lawrence Berkeley National Laboratory
(LBNL), under Contract DE-AC02-05CH11231.

\clearpage

\bibliography{IceLikeSurface.bib}

\clearpage



\section*{Supplementary material}

# Supplementary Information: Local ice-like structure at the liquid water surface

Nathan L. Odendahl[†,‡] and Phillip L. Geissler[*,†,‡]

†Department of Chemistry, University of California, Berkeley, California 94720, USA

‡Chemical Sciences Division, Lawrence Berkeley National Laboratory, Berkeley, California 94720, USA

E-mail: geissler@berkeley.edu

## The Joint Cosine Distribution

The depth-dependent orientational distribution $P(\cos\theta_{\text{OH}_1}, \cos\theta_{\text{OH}_2}; d)$ is nonuniform due not only to surface effects but also to a geometric bias that that exists even in an isotropic bulk environment. Here we work out the exact form of this geometric bias for a model water molecule whose rigid geometry imposes an OH bond distance $\ell_{\text{OH}}$, an HOH bond angle $\gamma$, and a corresponding HH distance $\ell_{\text{HH}} = 2\ell_{\text{OH}}\sin(\gamma/2)$. We determine $P_{\text{bulk}}(\cos\theta_{\text{OH}_1}, \cos\theta_{\text{OH}_2})$ by replacing the HOH bond angle constraint with a smooth potential energy $u(r)$, which allows small but finite deviations of the HH distance $r$ away from its ideal value $\ell_{\text{HH}}$. In the limit of a highly stiff potential, the constrained result should be recovered.

We first write $P_{\text{bulk}}(\cos\theta_{\text{OH}_1}, \cos\theta_{\text{OH}_2})$ as a marginal of the Boltzmann distribution,

$$P_{\text{bulk}}(\cos\theta_{\text{OH}_1}, \cos\theta_{\text{OH}_2}) = q^{-1}\int d\hat{\mathbf{R}}_1 \int d\hat{\mathbf{R}}_2 e^{-\beta u(r)}\delta(\cos\theta_{\text{OH}_1} - \hat{\mathbf{R}}_1\cdot\hat{\mathbf{z}})\delta(\cos\theta_{\text{OH}_2} - \hat{\mathbf{R}}_2\cdot\hat{\mathbf{z}}),$$

(1)

where $\hat{\mathbf{R}}_j$ is a unit vector along the $\mathrm{OH}_j$ bond, $\beta = 1/(k_\mathrm{B}T)$, $q$ is the normalizing partition function

$$q = \int d\hat{\mathbf{R}}_1 \int d\hat{\mathbf{R}}_2 e^{-\beta u(r)}, \qquad (2)$$

and $r = \ell_\mathrm{OH}|\hat{\mathbf{R}}_1 - \hat{\mathbf{R}}_2|$. Arbitrarily setting the azimuthal angle of $\mathrm{OH}_1$ at $\phi_1 = 0$, we obtain

$$P_\mathrm{bulk}(\cos\theta_{\mathrm{OH}_1}, \cos\theta_{\mathrm{OH}_2}) = \frac{2\pi}{q} \int_0^{2\pi} d\phi_2 e^{-\beta u(r)} \qquad (3)$$

Given an allowed set of angles $(\cos\theta_{\mathrm{OH}_1}, \cos\theta_{\mathrm{OH}_2})$, the fluctuating distance

$$r = \ell_\mathrm{OH}[2 - 2(\cos\theta_{\mathrm{OH}_1}\cos\theta_{\mathrm{OH}_2} + \sin\theta_{\mathrm{OH}_1}\sin\theta_{\mathrm{OH}_2}\cos\phi_2)]^{1/2} \qquad (4)$$

is equal to the ideal distance $\ell_\mathrm{HH}$ at two symmetry-related roots $\phi_2 = \pm\bar{\phi}_2$. (For a disallowed pair of OH orientations, no value of $\phi_2$ satisfies $r = \ell_\mathrm{HH}$.) For small deviations $\delta\phi_2$ about either of these roots,

$$r - \ell_\mathrm{HH} = \frac{\ell_\mathrm{OH}^2}{\ell_\mathrm{HH}} \sin\theta_{\mathrm{OH}_1}\sin\theta_{\mathrm{OH}_2}\sin\bar{\phi}_2 \delta\phi_2 + \mathcal{O}(\delta\phi_2^2) \qquad (5)$$

A change of integration variables in Eq. 3 then gives

$$P_\mathrm{bulk}(\cos\theta_{\mathrm{OH}_1}, \cos\theta_{\mathrm{OH}_2}) = \frac{4\pi}{q} \frac{\ell_\mathrm{HH}}{\ell_\mathrm{OH}^2} (\sin\theta_{\mathrm{OH}_1}\sin\theta_{\mathrm{OH}_2}\sin\bar{\phi}_2)^{-1} \int dr\, e^{-\beta u(r)}, \qquad (6)$$

where integration limits have been extended arbitrarily in light of the stiff restraining potential. This result can be simplified by noting that the roots of $r = \ell_\mathrm{HH}$ satisfy

$$\sin\bar{\phi}_2 = \left[1 - \left(\frac{1 - \cos\theta_{\mathrm{OH}_1}\cos\theta_{\mathrm{OH}_2} - \ell_\mathrm{HH}^2/(2\ell_\mathrm{OH}^2)}{\sin\theta_{\mathrm{OH}_1}\sin\theta_{\mathrm{OH}_2}}\right)^2\right]^{1/2} \qquad (7)$$

The partition function $q$ can be similarly evaluated, noting first that the integral over $\hat{\mathbf{R}}_2$

in Eq. 2 is insensitive to $\hat{\mathbf{R}}_1$ by symmetry. Choosing $\hat{\mathbf{R}}_1 = \hat{\mathbf{z}}$ for simplicity, we obtain

$$q = 8\pi^2 \int d\cos\theta_{\text{OH}_2} e^{-\beta u(r)}. \tag{8}$$

Expanding $r = 2\ell_{\text{OH}} \sin(\theta_{\text{OH}_2}/2)$ for $\cos\theta_{\text{OH}_2} \approx \cos\gamma$ and changing the variable of integration then gives

$$q = 8\pi^2 \frac{\ell_{\text{HH}}}{\ell_{\text{OH}}^2} \int dr\, e^{-\beta u(r)}, \tag{9}$$

where limits of integration have again been extended arbitrarily.

Combining Eqs. 6, 7, and 9, we finally obtain

$$P_{\text{bulk}}(\cos\theta_{\text{OH}_1}, \cos\theta_{\text{OH}_2}) =$$
$$(2\pi)^{-1} \left[ (1 - \cos^2\theta_{\text{OH}_1})(1 - \cos^2\theta_{\text{OH}_2}) - (1 - \cos\theta_{\text{OH}_1}\cos\theta_{\text{OH}_2} - b)^2 \right]^{-1/2}, \tag{10}$$

where

$$b = \frac{\ell_{\text{HH}}^2}{2\ell_{\text{OH}}^2} = 2\sin^2\left(\frac{\gamma}{2}\right). \tag{11}$$

For a perfectly tetrahedral molecular geometry, $\cos\gamma = -1/3$ and $b = \frac{4}{3}$. For TIP4P/Ice, the angle $\gamma = 104.52°$ is slightly less obtuse, giving $b \approx 1.25$.

Histograms measured in MD simulations agree well with Eq. 10. All cosine-cosine plots presented in the manuscript are normalized to the isotropic distribution measured in bulk MD, i.e., we plot $P(\cos\theta_{\text{OH}_1}, \cos\theta_{\text{OH}_2}; d)/P_{\text{bulk}}(\cos\theta_{\text{OH}_1}, \cos\theta_{\text{OH}_2})$.

## Extent of parallel tempering

The parallel tempered simulations exchange over a range of temperatures from 17 K to 287 K. Fig. SI.1 presents illustrative temperature histories of four trajectories. Although significant exchange occurs at all temperatures, simulations that began cold tend to stay cold while temperatures that began warm tend to stay warm, with a soft boundary near 160 K. Due to the low rate of exchange, the data from temperatures below 137 K are unlikely to be well sampled, and so only temperatures at or above 137 K are used to form conclusions in this work.

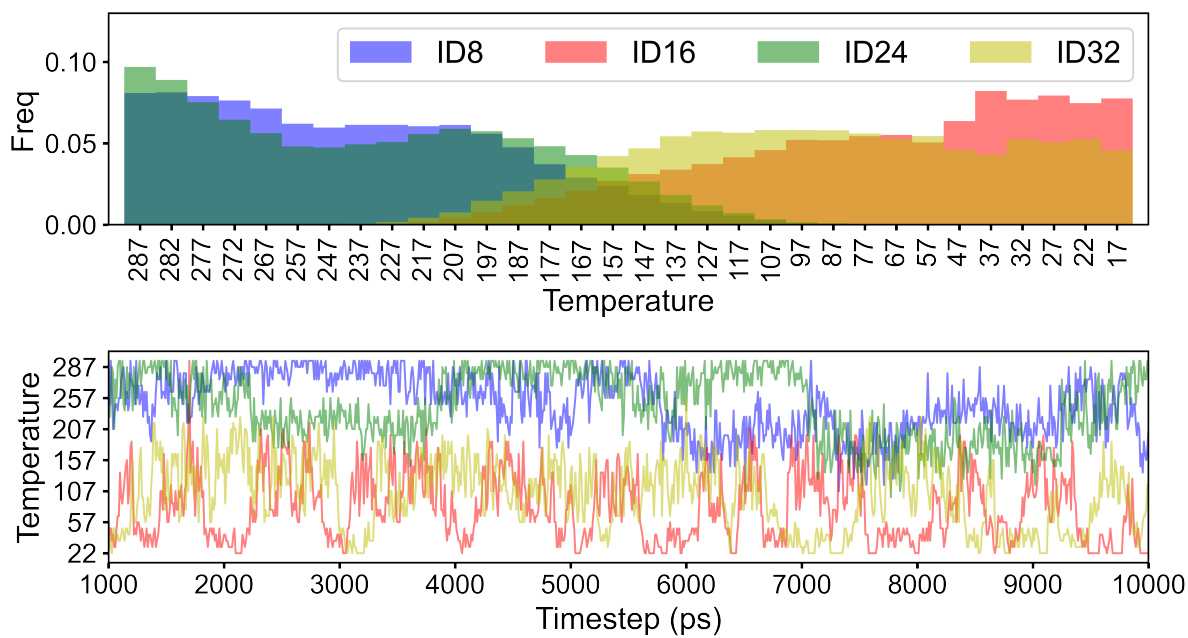

Figure 1: Representative sample of the temperature exchange of four simulations during parallel tempered simulations of ice as a histogram (top) or by timestep (bottom).

## Layering in the ice crystal

The density profiles of the primary prismatic face (Fig. SI.2) and secondary prismatic face (Fig. SI.3) exhibit clear layering that can also be related to the liquid water surface. Some layering at intervals of approximately 3 Å follows from the average bond length of water. However, the other two ice crystal faces do not correspond to the liquid water surface density as closely as the basal face of ice.

The density layering at the ice basal surface diminishes smoothly as temperature increases (Fig. SI.4), with L1A in particular losing clear structure near melting.

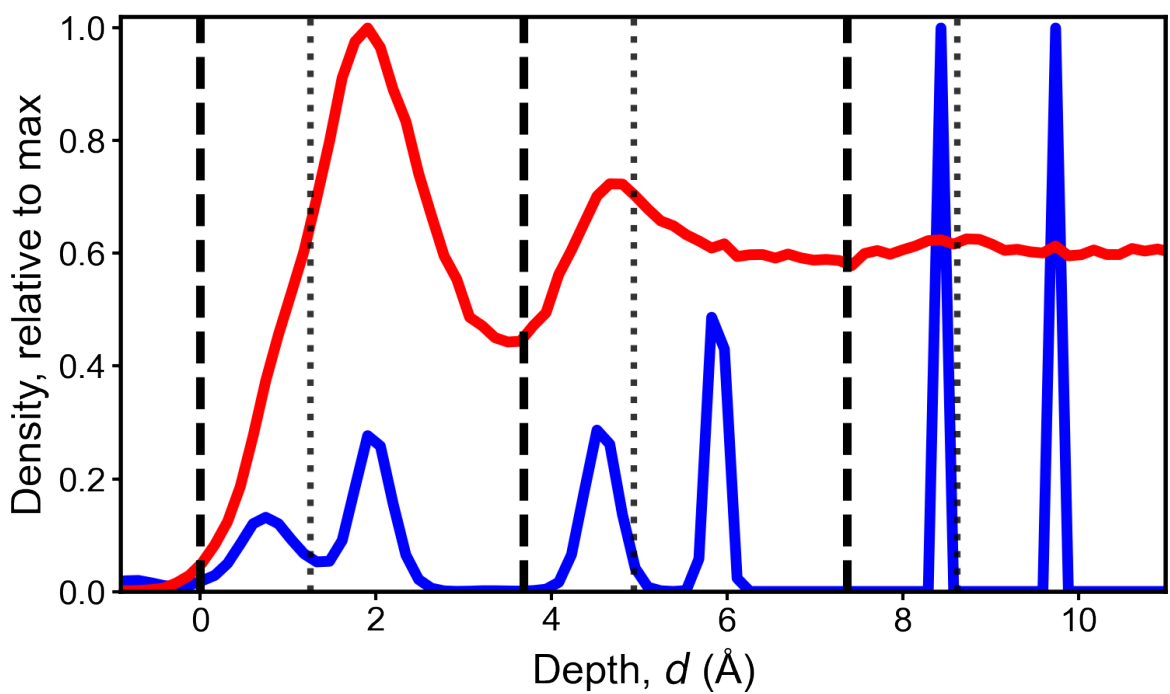

Figure 2: Comparison of the density profile of the primary prismatic face (blue) with the liquid water surface (red).

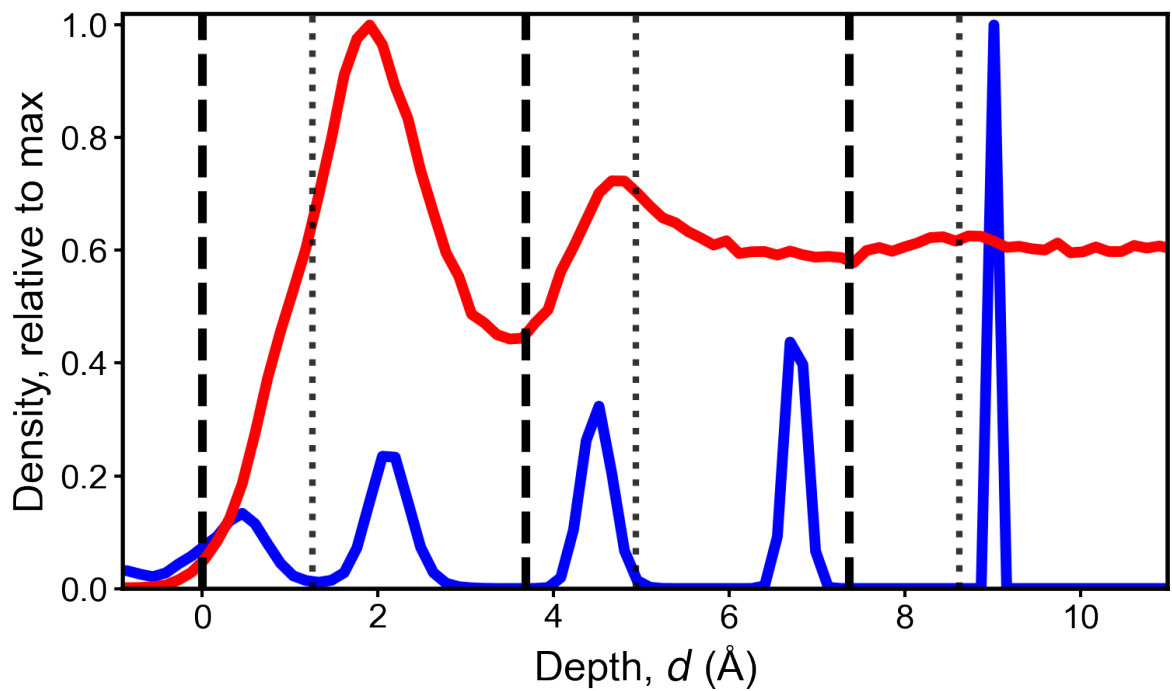

Figure 3: Comparison of the density profile of the secondary prismatic face (blue) with the liquid water surface (red).

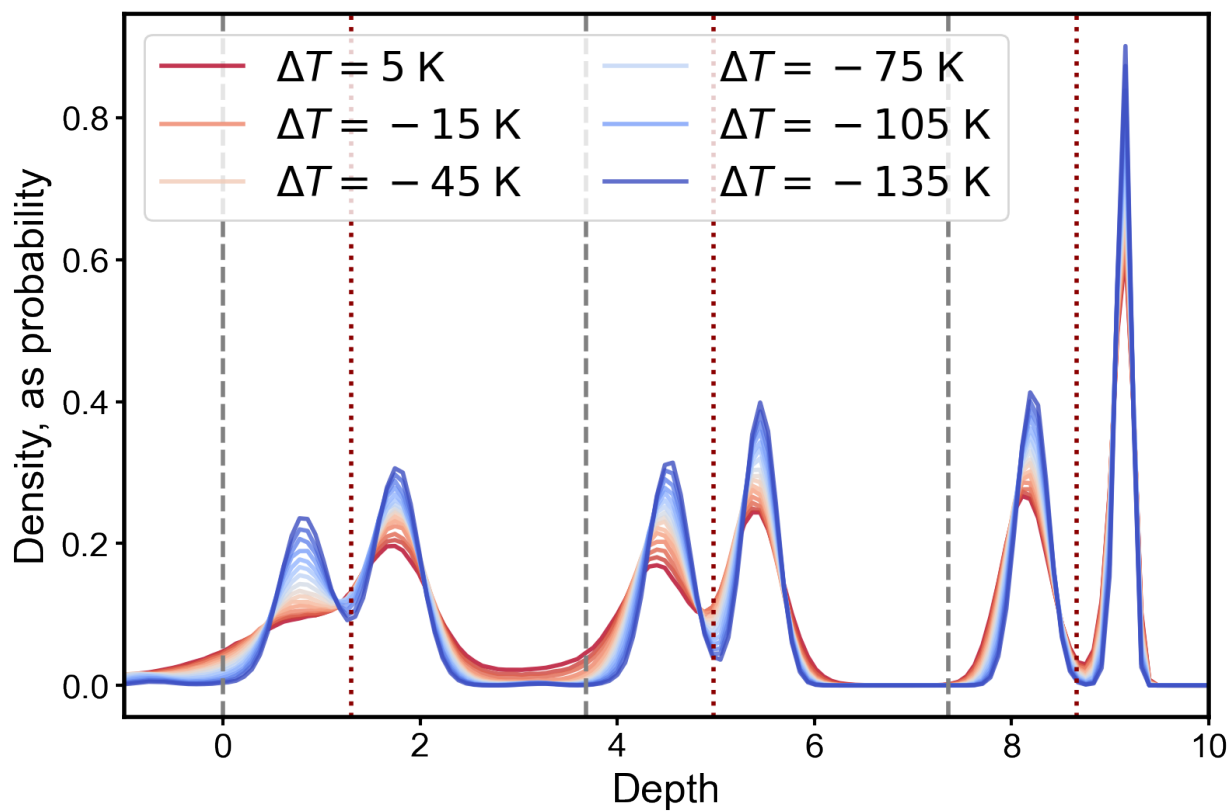

Figure 4: Summary of the change in density profile with increasing temperature according to parallel tempered simulations. Layers L1 to L3 are displayed here and L4 is held stationary throughout each simulation, which partially causes the sharp peak in L3B.

# Orientations in the primary and secondary ice faces

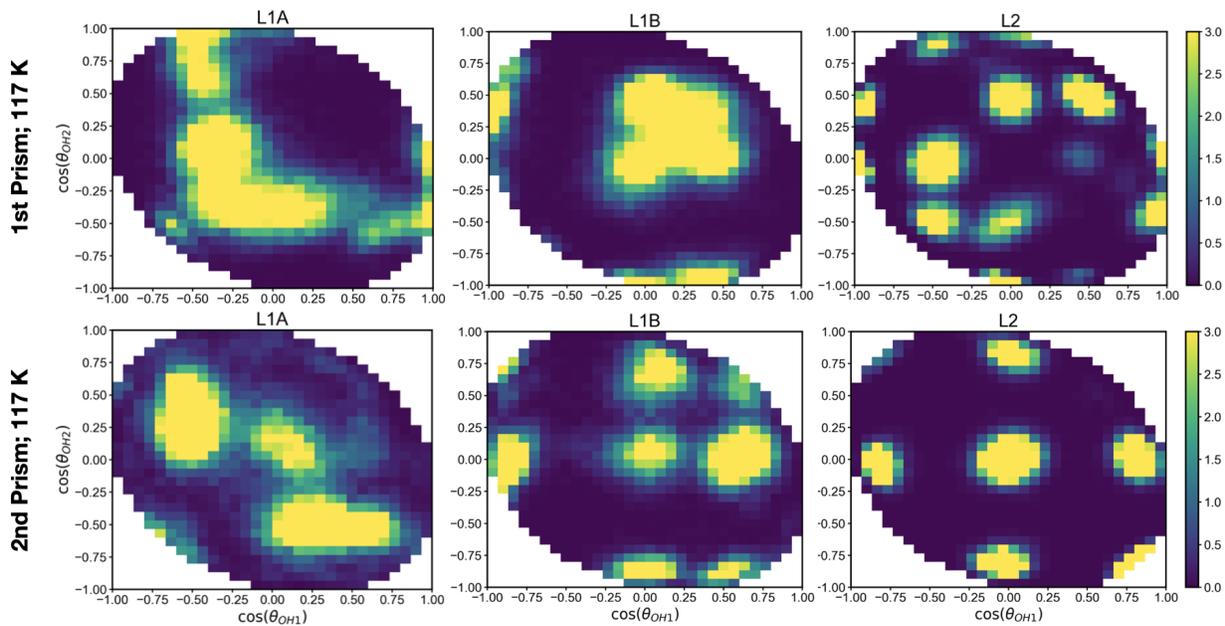

Figure 5: Joint OH orientational measurements of the primary prismatic (top) and secondary prismatic (bottom) faces of ice.

Water molecular orientations measured at the primary and secondary ice faces (Fig. SI.5) are substantially different from that measured at the water surface. In contrast (as discussed in the main text) the basal face of ice exhibits molecular orientations remarkably similar to the liquid water surface.

## Alternative definitions of L1$^\parallel$

While a parallel layer at the water surface has been observed elsewhere,[1-3] no consistent boundary is used to define the parallel region. "Parallel" in this region means that water molecules tend to point both hydrogen mainly in the plane of the interface, which is usually defined using a measurement of the angle between the surface normal $\hat{n}$ and with OH$_i$ or the dipole $\mu$.

In crystalline ice, L1A $\cos\theta_{\text{OH}_i}$ is equally centered around either 1.0 or -0.33, and the corresponding L1A dipole $\cos\theta_\mu$ is centered around either 0.6 or -0.33. The peak positions swap in L1B, with peaks in $\cos\theta_{\text{OH}_i}$ at combinations of -1.0 and +0.33. In defining the parallel layer, angular cutoffs need to avoid the known L1A and L1B peaks around $\cos\theta_{\text{OH}_i} \pm 0.33$. A window $\cos\theta^\parallel \in [-0.25, 0.25]$ corresponds to an arc length of about 30°.

At the liquid surface, $\cos\theta_\mu \in [-0.25, 0.25]$ suggests about 32% of water molecules in L1 are parallel, but if both $\theta_{\text{OH}_i} \in [-0.25, 0.25]$ then only about 8.5% of L1 waters qualify. The surface population of parallel water molecules ranging 8.5-32% is a minority of L1 waters, but doubles an equivalent measurement in bulk. A narrow slice of increased parallel character certainly exists near the L1A/B boundary.

The L1$^\parallel$ region also exhibits a high degree of interconnected hydrogen bonds.[2-4] For a water molecule sitting in L1$^\parallel$, first-neighbor hydrogen bonds can connect to other water molecules within L1$^\parallel$ (intra), above L1$^\parallel$ (upper), or below L1$^\parallel$ (lower). Serva et al.[4] measured that connections between L1$^\parallel$ molecules (intra) occur at nearly twice the rate as in bulk liquid. Setting L1$^\parallel$ boundaries to $1.0 < d < 2.8$ Å maximizes "intra" bonding by capturing the lower portion of L1A and nearly all of L1B. The parallel character is likely caused by a mixture of the underlying ice-like structure, the low density of bonding neighbors within L1A, and a reversion to isotropic orientations at higher temperature. The abnormal increase in "intra" hydrogen bonding in L1B can be rationalized as a tendency of L1A molecules to cross into the L1B region (main text Fig. 4 and Fig. SI.6), and is still consistent with the quasi-ice picture of the liquid water surface. Table SI.1 summarizes upper/intra/lower

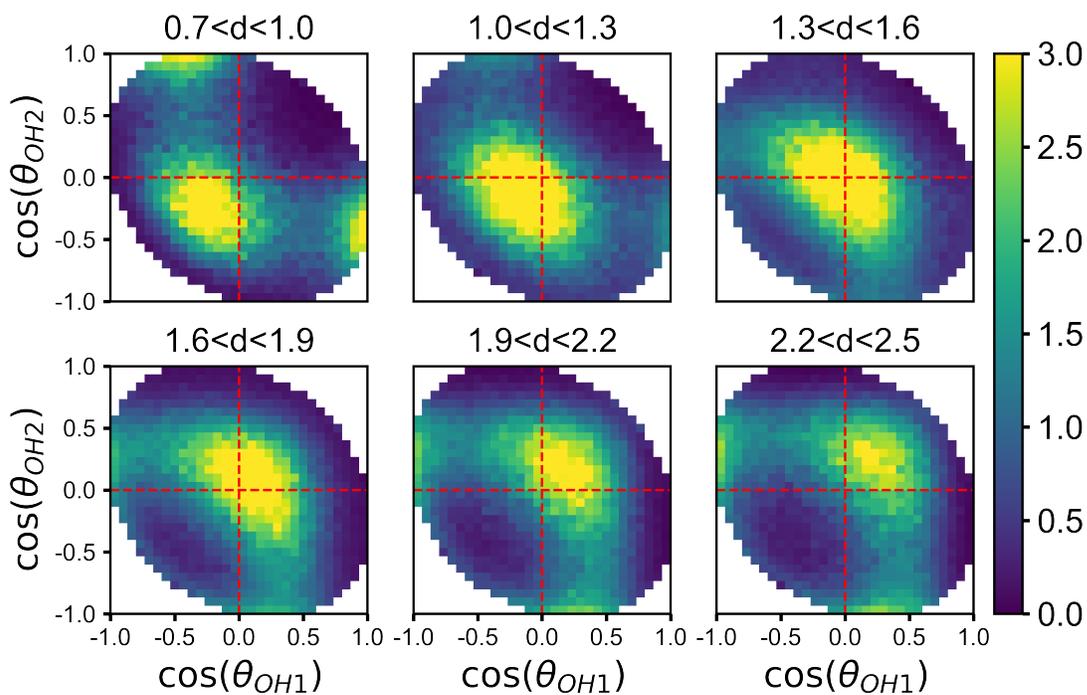

Figure 6: Progression of water molecule orientations by 0.3 Å slice of the liquid water surface. Grid lines at $\cos\theta_{\text{OH}_i} = 0$ are emphasized in red.

hydrogen bonding measured at the layer boundaries.

Fig. SI.6 of the progression of orientations through L1 reveals three main regimes: L1A, L1$^\parallel$, and L1B. The smooth transition suggests that L1A molecules shift to and from L1B by pointing both OH in the plane of the interface and passing through L1$^\parallel$. While there is a characteristic parallel region between $d = 1.0$ to $d = 1.6$, the total density of molecules in that region is small. Overall, the majority of liquid L1 water molecules are primarily classified as L1A or L1B, and much of the parallel population can be rationalized as a transition between L1A and L1B positions.

Table 1: Readout of average upper/intra/lower hydrogen bonding behavior of various layer-slices of TIP4P/Ice for liquid (298 K) and ice (137 K, L4 held stationary). The definition of a hydrogen bond follows the procedure of White et al.[5]

| FOR TIP4P/ICE @ 298 K | WIDTH | UPPER / INTRA / LOWER | TOTAL HBONDS |
|---|---|---|---|
| BULK | 1.3 | 1.35 / 0.82 / 1.35 | 3.52 |
| BULK | 1.8 | 1.20 / 1.13 / 1.19 | 3.52 |
| BULK | 2.38 | 1.02 / 1.49 / 1.01 | 3.52 |
| BULK | 3.68 | 0.68 / 2.17 / 0.67 | 3.52 |
| L1 | 3.68 | 0.01 / 2.43 / 0.72 | 3.16 |
| L1 ÷ BULK | 3.68 | 0.01 / 1.12 / 1.07 | 0.90 |
| L1A | 1.3 | 0.01 / 0.43 / 1.92 | 2.37 |
| L1A ÷ BULK | 1.3 | 0.01 / 0.53 / 1.43 | 0.67 |
| L1$^{\parallel}$ | 1.8 | 0.32 / 1.81 / 1.07 | 3.20 |
| L1$^{\parallel}$ ÷ BULK | 1.8 | 0.27 / 1.60 / 0.90 | 0.91 |
| L1B | 2.38 | 0.47 / 1.99 / 0.89 | 3.36 |
| L1B ÷ BULK | 2.38 | 0.47 / 1.34 / 0.88 | 0.95 |
| L2 | 3.68 | 0.67 / 2.21 / 0.66 | 3.53 |
| L2A | 1.3 | 1.30 / 0.86 / 1.38 | 3.54 |
| L2B | 2.38 | 1.05 / 1.50 / 0.99 | 3.53 |
| FOR TIP4P/ICE @ 137 K | WIDTH | UPPER / INTRA / LOWER | TOTAL HBONDS |
| L1 | 3.68 | 0.03 / 2.99 / 0.52 | 3.54 |
| L1A | 1.3 | 0.06 / 0.11 / 2.90 | 3.07 |
| L1B | 2.38 | 2.40 / 0.59 / 0.93 | 3.92 |
| L2 | 3.68 | 0.51 / 2.99 / 0.50 | 4.00 |
| L2A | 1.3 | 1.01 / 0.01 / 2.98 | 4.01 |
| L2B | 2.38 | 2.93 / 0.09 / 0.99 | 4.00 |

Table 2: Density of water molecules in each surface layer, relative to L4 (effectively bulk liquid).

| FOR TIP4P/ICE @ 298 K | Layer molecule count relative to count in L4 |
|---|---|
| L0 ÷ L4 | 0.005 |
| L1 ÷ L4 | 0.94 |
| L1A÷L4 | 0.19 |
| L1B÷L4 | 0.76 |
| L2÷L4 | 1.02 |
| L2A÷L4 | 0.36 |
| L2B÷L4 | 0.66 |
| L3÷L4 | 0.99 |
| L3A÷L4 | 0.36 |
| L3B÷L4 | 0.64 |

## Minimal model of the surface dipole: $m_{\text{est}}$

To model the average dipole profile for finite-temperature ice, we classify each water molecule as either "inter" or "intra", according to the locations of its hydrogen bonding partners. An "intra" molecule directs both OH groups within its own layer. An "inter" molecule directs one OH group within the layer and the other towards a different layer. In L$X$A an "inter" molecule's dipole points strongly up along $\hat{\boldsymbol{n}}$ ($\cos\theta_{\text{OH}} \approx 1$, $\cos\theta_\mu \approx 0.58$), while an "intra" molecule's dipole points more weakly down ($\cos\theta_{\text{OH}} = \cos\theta_\mu \approx -1/3$). In a symmetric way, within L$X$B an "inter" molecule's dipole points strongly down $\cos\theta_{\text{OH}} \approx -1$, while an "intra" molecule's dipole points more weakly up ($\cos\theta_{\text{OH}} \approx +1/3$).

Less than approximately one in three of the molecules in a given sublayer of ice should be of type "inter", lest the system as a whole acquire a macroscopically large dipole moment. For a tetrahedral bonding geometry, this constraint implies that each sublayer is non-dipolar on average. Systematic displacements of "inter" and "intra" molecules in opposite directions, however, can yield a nonzero dipole profile $\langle m_{\text{n}}(d)\rangle \neq 0$ while still respecting the requirement that $\langle m_{\text{n}}(d)\rangle \rho(d)$ integrates to zero over the sublayer. For layers $X > 1$, our simulation results thus suggest that "inter" molecules in L$X$A shift downwards (towards bulk) while L$X$A "intra" molecules shift upwards (towards the surface). Symmetrically in L$X$B, "inter" molecules instead shift upwards, and "intra" molecules downwards. The outermost surface causes a unique exception, where L1A reverses the regular dipole pattern and "inter" molecules (i.e., dangling OH groups) shifting upwards.

To test this understanding of the ice dipole profile, we attempt to reconstruct it as a weighted sum of characteristic dipole moments $\bar{m}_i^\alpha$, where $i$ indicates "inter" or "intra", and $\alpha$ indicates a specific sublayer. This estimate,

$$m_{\text{est}}(d) = \sum_i \bar{m}_i^{\alpha(d)} \rho_i(d), \tag{12}$$

is determined by computing $\bar{m}_i^\alpha$ as an average dipole over the corresponding molecular pop-

ulation in computer simulations. In Eq. SI.12, $\rho_{\text{inter}}(d)$ and $\rho_{\text{intra}}(d)$ are density profiles for "inter" and "intra" populations, and $\alpha(d)$ returns the sublayer corresponding to depth $d$. To capture this relation at warmer temperatures when near-ice structure is expected, $|\cos\theta_{\text{OH1}} - \cos\theta_{\text{OH2}}| < 0.3$ indicates that both OH vectors are likely "intra" while the remaining $|\cos\theta_{\text{OH1}} - \cos\theta_{\text{OH2}}| > 0.3$ would likely point "inter". The comparison in Fig. 5 of the main text suggests that this estimate, and its underlying physical picture, are generally sound.

We make four observations about the average surface dipole of rigid point-charge water. First, that there is an oscillating bilayer-by-bilayer dipole pattern at the ice surface at non-zero temperatures. Second, that the bilayer in ice exhibits sub-bilayer structure due to subtle asymmetries in donor and acceptor geometries. Third, that the dipole within L1 is similar between ice and liquid water, with L1A highly similar, and that this is likely due to a shared layering-with-defects structure. And fourth, that the sub-bilayer structure reverses within L1A, consistent with the picture that the "intra" molecules dip well below their regular height to cross into L1B, while L1A "inter" molecules (those with a dangling hydrogen bond) sit abnormally high.



\end{document}